\documentclass{aa}  

\usepackage{graphicx}
\usepackage{siunitx}
\usepackage{xcolor}
\usepackage{multirow}
\usepackage{longtable}
\usepackage{booktabs}
\usepackage{array}
\usepackage{amsmath}
\usepackage{xspace}
\usepackage{txfonts}
\usepackage{hyperref}
\usepackage{natbib}

\newcommand{\thethree}{{\sc The300}\xspace}
\newcommand{\theth}{{\sc The Three Hundred}\xspace}
\newcommand{\planck}{\textit{Planck}\xspace}
\newcommand{\pszcosmo}{PSZ2\_\emph{cosmo}\xspace}

\begin{document} 

   \title{Inference of morphology and dynamical state of nearby \planck-SZ galaxy clusters with Zernike polynomials}

   \author{V. Capalbo \inst{1}  
          \and
          M. De Petris \inst{1}
          \and
          A. Ferragamo \inst{1,2}
          \and
          W. Cui \inst{3,4,5}
          \and
          F. Ruppin \inst {6}
          \and
          G. Yepes \inst {4,5}
          }

   \institute{Dipartimento di Fisica, Sapienza Università di Roma, Piazzale Aldo Moro 5, I-00185 Roma, Italy\\
              \email{valentina.capalbo@uniroma1.it}
         \and
         Physics Department “Ettore Pancini”, Università degli studi di Napoli “Federico II”, Via Cintia 21, I-80126 Napoli, Italy
         \and
         Institute for Astronomy, University of Edinburgh, Royal Observatory, Edinburgh EH9 3HJ, UK
         \and
         Departamento de Física Teórica, Módulo 8, Facultad de Ciencias, Universidad Autónoma de Madrid, E-28049 Madrid, Spain
         \and
         Centro de Investigación Avanzada en Física Fundamental (CIAFF), Facultad de Ciencias, Universidad Autónoma de Madrid, E-28049 Madrid, Spain
         \and
         University of Lyon, UCB Lyon 1, CNRS/IN2P3, IP2I Lyon, France}

   \date{Received ...; accepted ...}

  \abstract
   {}
   {We analyse the maps of the Sunyaev-Zel'dovich (SZ) signal of local galaxy clusters ($z<0.1$) observed by the \planck satellite in order to classify their dynamical state through morphological features.}
   {To study the morphology of the cluster maps, we apply a method recently employed on mock SZ images generated from hydrodynamical simulated galaxy clusters in \theth (\thethree) project. Here, we report the first application on real data. The method consists in modelling the images with a set of orthogonal functions defined on circular apertures, the Zernike polynomials. From the fit we compute a single parameter, $\mathcal{C}$, that quantifies the morphological features present in each image. The link between the morphology of 2D images and the dynamical state of the galaxy clusters is well known, even if not obvious. We use mock \planck-like Compton parameter maps generated for \thethree clusters to validate our morphological analysis. These clusters, in fact, are properly classified for their dynamical state with the relaxation parameter, $\chi$, by exploiting 3D information from simulations.}
   {We find a mild linear correlation of $\sim 38\%$ between $\mathcal{C}$ and $\chi$ for \thethree clusters, mainly affected by the noise present in the maps. 
   In order to obtain a proper dynamical-state classification for the \planck clusters, we exploit the conversion from the $\mathcal{C}$ parameter derived in each \planck map in $\chi$.
   A fraction of the order of $63\%$ of relaxed clusters is estimated in the selected \planck sample.
   Our classification is then compared with those of previous works that have attempted to evaluate, with different indicators and/or other wavelengths, the dynamical state of the same \planck objects. The agreement with the other works is larger than $58\%$.}
   {}


   \maketitle

\section{Introduction}
\label{sec:intro}
Galaxy clusters constitute the final stage of the hierarchical structure formation process \citep[see e.g.][]{Voit2005} and their census in terms of mass and redshift distribution is one of the ways to put constraints on cosmological parameters \citep[see e.g.][]{Planck2014,PSZ_cosmo1,PSZ_cosmo2}.
The accurate estimation of the mass is, of course, the main ingredient in cosmological constraints from galaxy cluster abundances.
Different approaches are explored to infer the mass, not being a direct observable, but with simplified assumptions.
For example, assuming a self-similar model \citep{Kaiser1986}, several scaling relations are defined to relate the mass to multi-wavelengths observables and are used to constrain the mass when dealing with large samples of systems \citep[see e.g.][]{Giodini2013,Pratt2019}. 
However, some simplified hypotheses are used, such as the hydrostatic equilibrium based on spherical distribution for dark matter and baryon content \citep[see e.g.][]{Kravtsov_Borgani2012}. It is well known that galaxy clusters during their evolution can experience states of non-equilibrium, for example during merger events. This condition may introduce biases in mass computation \citep{Planck2014}. Classifying clusters according to their dynamical state can help account uncertainties in these methods.
From an observational point of view, the analysis of the morphological appearance of cluster images generated at different wavelengths is the cheapest way to attempt a dynamical-state assessment. The presence of irregularities at different scales in the images can in fact be sign of dynamical activities in the systems not related to the conditions of equilibrium described before. The first approach to the morphological analysis was carried out through visual classifications of cluster images from X-ray observations \citep{Jones_Forman1992}. A method that, however, is subjective and difficult to be applied to large samples. Different techniques were then developed to speed up and make these analyses more robust, e.g. defining several parameters that can quantify morphological differences between images. Some of the most commonly used are, for example, 
the centroid shift \citep{Mohr1993,OHara2006}, the power ratios \citep{Buote1995}, the asymmetry \citep{Schade1995}, the cuspiness of the gas density profile \citep{Vikhlinin2007}, the concentration ratio \citep{Santos2008}, the central gas density \citep{Hudson2010}, the Gini coefficient \citep{Parekh2015}.
They are used singularly or combined together \citep[see e.g.][]{Rasia2013,Mantz2015,Lovisari2017,Cialone2018,Bartalucci2019,DeLuca2021,Campitiello2022}, taking into account the cluster apertures to analyse, for example being sensitive to the cores or to characteristic radii (e.g. $R_{500}$\footnote{$R_{500}$ is the radius enclosing an overdensity of 500 times the critical density of the Universe at cluster's redshift. A similar definition applies to the subscript 200.}, $R_{200}$ and so on). Several approaches are employed considering the different cluster components, i.e. galaxies and gas \citep[see e.g.][]{Ribeiro2013,Wen2013,Lopes2018,Cerini2023}. For example, by using optical and X-ray data, it is possible to measure the offset between the position of the bright central galaxy (BCG) and the X-ray peak or centroid \citep[see e.g.][]{Jones_Forman1984,Lavoie2016,Lopes2018,DeLuca2021} or the other bright member galaxies \citep{Casas2024}. Also, the magnitude difference between the first and second BCGs can be employed as a dynamical state indicator, as shown in \citet{Lopes2018}.
To date, a large literature is available about the definition and the application of morphological parameters both on X-ray and optical observations. Several works also applied these methods on simulated data \citep[see e.g.][]{Cao2021,DeLuca2021,Li2022,Seppi2023}. 
Recently, the morphological analysis of galaxy clusters has been also extended to simulations of millimetre observations through the Sunyaev-Zel'dovich (SZ) effect \citep[][C21 hereafter]{Cialone2018,DeLuca2021,Capalbo2021}. The SZ effect, in fact, opened a new window to produce large samples of galaxy clusters.
It is a spectral distortion of the cosmic microwave background (CMB) radiation caused by the interaction, via inverse Compton scattering, of the photons with the energetic free electrons in the intracluster medium (ICM) \citep{SZ1972}.
In particular, the thermal component of the SZ effect (tSZ), that depends on the thermal random motion of the electrons, causes a boost in the energy of the CMB photons resulting in an increase of the CMB brightness at frequencies $\gtrsim$ 217 GHz in the direction of the clusters \citep[see e.g][for a review]{Mroczkowski2019}. The intensity of the effect is quantified with the Compton parameter, $y$:
\begin{equation}
    \label{eq:y}
    y (\hat{n}) = \frac{\sigma_T}{m_e c^2} \int P_e(\hat{n},l) dl
\end{equation}
where $\sigma_T$ is the Thomson scattering cross-section, $m_e$ is the electron rest mass, $c$ is the speed of light, $P_e$ is the electron pressure and the integration is done along the line of sight $l$ in the direction $\hat{n}$.
The advantages of tSZ surveys with respect to X-ray ones are two: the independence of the spectral distortion from the redshift and the linear dependence of $y$ from the electron pressure (at least in the non-relativistic approximation). These make the tSZ effect well suitable to observe the most distant clusters and construct samples that are, in principle, mass limited and representative of the underlying population.
Several instruments have been devoted to observe galaxy clusters through the tSZ effect. The \planck satellite made it possible to produce full-sky Compton parameter maps (i.e. $y$-maps) \citep{Planck_y_maps} and a catalogue of tSZ sources which, in its last release from the full-mission data \citep{PSZ2} contains 1653 detections with 1203 confirmed clusters from ancillary data. The South Pole Telescope (SPT) was the first ground-based instrument suitable to observe distant galaxy clusters through the tSZ effect \citep{Staniszewski2009}. Since then, a large number of clusters have been detected in several SPT surveys \citep{Bleem2015,Huang2020,Bleem2020}, with also the recent release of a $y$-map covering $\sim2500$ deg$^2$ of the southern sky at $1.25^{\prime}$ resolution \citet{Bleem2022}.
The redshift coverage of tSZ-selected galaxy clusters has been improved with the Atacama Cosmology Telescope (ACT) that, as in the case of SPT, provided several samples of clusters during the years \citep{Hilton2018,Hilton2021}. The last catalogue derived from this survey \citep{Hilton2021} contains > 4000 optically confirmed clusters, with 222 at $z>1$ and a total of 868 systems that are new discoveries, while a new $y$-map at $1.6^{\prime}$ resolution and covering $\sim 13 000$ deg$^2$ of the sky has been recently released \citep{Coulton2024}.

In this work we use the all-sky $y$-map provided by the \planck satellite to study the morphology of nearby ($z<0.1$) galaxy clusters.
We apply the method developed in C21 which consists in modelling the maps analytically by using the Zernike polynomials (ZPs hereafter), in order to quantify their different morphologies. This approach was tested in C21 on a set of high-resolution $y$-maps generated from synthetic clusters in {\sc{The Three Hundred}} project\footnote{https://the300- project.org} \citep[hereafter \thethree;][]{Cui2018}. Modelling with the ZPs has proven to be a suitable way to analyse a wide variety of images, providing the chance of exploring different spatial scales in the maps simply by changing the number of ZPs used, and being able to reveal the main features for a good morphological classification. In C21 a single parameter, $\mathcal{C}$, was introduced to collect the capability of ZPs to describe the different morphologies in the $y$-maps. $\mathcal{C}$ showed a good correlation with most common morphological parameters used in the literature and also with a combination of some proper dynamical state indicators available, in that case, from simulations \citep{DeLuca2021}. Motivated by those results, here we apply the Zernike modelling for the first time on real data.
The aim of this work is to deduce a morphological classification for the \planck-SZ clusters using the $y$-maps and to verify if this analysis can allow to infer their dynamical state. To reach this goal, we use a set of mock \planck-like $y$-maps realized from \thethree clusters of which the dynamical state can be estimated a priori from 3D data. Hence, we use them to test if and how our morphological analysis of $y$-maps can be related to a proper dynamical classification of the clusters in the \planck sample. The link between the morphology and the dynamical state is, in fact, not obvious. Some limitations due, e.g., to projection effects or limited angular resolutions in the maps can reduce the capability to reveal characteristic features indicative of the dynamical state.

The study of the dynamical state of samples of \planck clusters was already conducted in several works \citep{Rossetti2016,Rossetti2017,Lovisari2017,Andrade-Santos2017,Lopes2018,Campitiello2022} by using X-ray or optical observations. In this work, we would like to provide a more complete view on this question by exploiting also the available millimetre data. In the end, we compare our classification with the ones reported in literature. 

The paper is structured as follows: in Section~\ref{sec:data_sets} we introduce the data sets; in Section~\ref{sec:methods} we recall the ZPs definition and the method employed to model the $y$-maps; in Section~\ref{sec:results} we report the results of the application of ZPs as morphological indicators and the calibration to infer the dynamical state; in Section~\ref{sec:comparisons} a comparison with literature data is analysed while our conclusions are summarized in Section~\ref{sec:conclusions}. Throughout this work, we assume a $\Lambda$CDM cosmology with $H_0=67.8$ km s$^{-1}$ Mpc$^{-1}$, $\Omega_m=0.307$ and $\Omega_{\Lambda}=0.048$ \citep{Planck2016}.

\section{Data sets}
\label{sec:data_sets}

\subsection{The \planck-SZ clusters sample and \textit{y}-maps}
\label{sec:PSZ2_cosmo}
\begin{figure}
    \centering
	\includegraphics[width=\columnwidth]{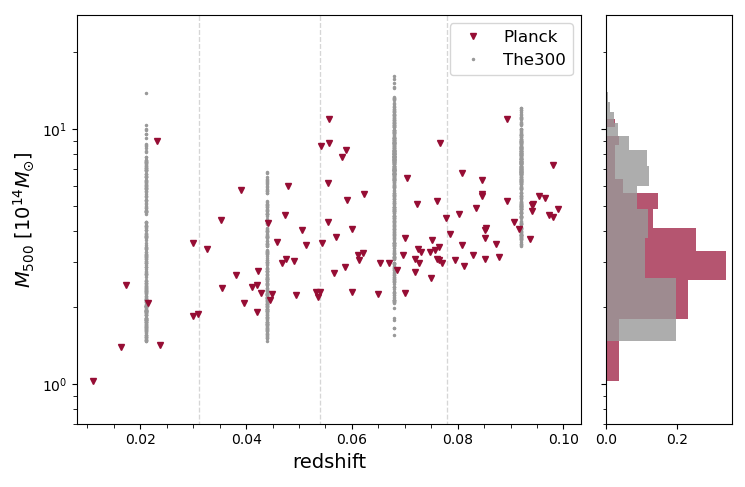}
    \caption{Distribution in the $M_{500}-z$ plane of 109 \planck-SZ clusters (red triangles) selected for the morphological analysis (see Section~\ref{sec:PSZ2_cosmo}). The thin dashed vertical lines mark the redshift bins used for the comparison with the synthetic clusters (grey dots) from {\sc The300} project (see Section~\ref{sec:The300}). In the right panel there are the normalized distributions of the masses for the \planck (in red) and \thethree (in grey) samples.}
    \label{fig:M_z}
\end{figure}

We select a sample of galaxy clusters from the second \planck catalogue of Sunyaev-Zel'dovich sources \citep[PSZ2,][]{PSZ2}. Specifically, we use one of the two subsamples extracted from the PSZ2 for cosmology studies \citep{PSZ_cosmo1, PSZ_cosmo2}, i.e. the intersection catalogue (\pszcosmo, hereafter). This catalogue is constructed by considering the detections made by all the three different algorithms used for the PSZ2, i.e. two Matched Multi-Filters techniques, MMF1 and MMF3 \citep[][]{Melin2006,Melin2012}, and the PowellSnakes (PwS) method \citep[][]{Carvalho2009,Carvalho2012}. All clusters show a signal-to-noise (S/N) $>6$. The catalogue covers 65\% of the sky, excluding the Galactic plane and bright point sources that can lead to spurious detections. We exclude clusters with potential infra-red contamination, indicated in the catalogue with a proper flag ([IR\_FLAG]=1). We also exclude the clusters without redshift association.
As the second step in defining the sample, we use the mass estimate, $M_{500}$, provided in the catalogue, to compute the angular radius $\theta_{500}$ of the clusters. \citet{PSZ2} specifies that the mass value is derived using the scaling relation $Y_{500}-M_{500}$ from \citet{Planck2014}, in which a mean bias of $(1-b)=0.8$ is assumed between the true mass and the hydrostatic mass. Therefore, we correct the mass values in the catalogue for the bias above, in order to obtain an estimate of the true mass of the clusters.
The simulated cluster masses are selected to be consistent with them, as described in Section~\ref{sec:The300}.
From $M_{500}=500\times (4/3)\pi R_{500}^3 \rho_c(z)$, where $\rho_c(z)$ is the critical density of the Universe at redshift $z$, we compute the cluster radius $R_{500}$, hence the respective angular size $\theta_{500}=R_{500}/D_A(z)$, where $D_A(z)$ is the angular diameter distance. Since the $y$-maps that we use have a resolution of $10^\prime$, as described in the next section, we select only the resolved objects at $z<0.1$, namely the clusters with $\theta_{500}\geqslant10^\prime$.
Our final sample is composed by 109 clusters. In Fig.~\ref{fig:M_z} we show, in red, the distribution of the clusters in the $M_{500}-z$ plane.

The public available all-sky $y$-maps, described in \citet{Planck_y_maps}, are constructed by combining the \planck single-channel maps convolved to a common resolution of 10$^\prime$ and by using two different component separation algorithms: MILCA \citep[Modified Internal Linear Combination Algorithm,][]{Hurier2013} and NILC \citep[Needlet Independent Linear Combination,][]{Remazeilles2011}. In this work we limit the analysis to the MILCA $y$-map. The all-sky $y$-maps are released in {\small HEALPix} \footnote{http://healpix.sf.net} format \citep[][]{Gorski2005}, with $N_\textit{side}$=2048 and pixel size of $1.7^\prime$. We use {\small healpy} \citep{Zonca2019}, a specific Python package to process {\small HEALPix} maps, to extract the $y$-maps of each cluster.
These latter are square cut-outs of side-length equal to $2\theta_{500}$, drawn as gnomonic projections from the all-sky $y$-map and with the same pixel resolution of the {\small HEALPix} map. Each cut-out is centred on the cluster coordinates reported in the PSZ2 catalogue. On these maps we overlap a circular aperture of radius $\theta_{500}$ to define the domain in which to apply the Zernike modelling. We also normalize the signal distribution to the maximum value of $y$ within the aperture.

\subsection{ \theth project}
\label{sec:The300} 
\thethree project \citep{Cui2018} is a large catalogue of hydrodynamically simulated galaxy clusters. The basis of the sample is composed by 324 large spherical regions, with radius of $15\,h^{-1}$Mpc, selected within the DM-only MultiDark simulation \citep[MDPL2,][]{Klypin2016}. These are the regions with most massive objects (virial mass $\gtrsim 8 \times 10^{14} h^{-1} M_{\odot}$ at $z=0$) at their centre, that are re-simulated with different baryonic physics models.
We use the catalogue of galaxy clusters generated with the Smooth-Particle-Hydrodynamics (SPH) code {\small GADGET-X} \citep{Beck2016}, that includes black hole and active galactic nuclei feedback.
128 redshift snapshots, from $z=0$ to $z=17$, are generated for each re-simulated region and the halo finder {\small AHF} \citep{AHF} is used to identify the virialized structures.
Here, we select 4 redshift snapshots, at $z=0.022, 0.044, 0.068$ and 0.092, to cover the redshift range of the \pszcosmo sample. The mass distribution along the different redshifts is plotted in Fig.~\ref{fig:M_z} in grey.

\subsubsection{Mock \planck-like \textit{y}-maps}
\label{sec:The300_y_maps} 
For \thethree galaxy clusters, mock $y$-maps are generated by using the {\small PyMSZ} code\footnote{https://github.com/weiguangcui/pymsz}, as described in \citet{Cui2018}. The procedure consists in discretizing the Eq.~\ref{eq:y} as follows:
\begin{equation}
    \label{eq:discrete_y}
    y \simeq \frac{\sigma_{T}k_{B}}{m_{e}c^{2}dA}\sum_{i}T_{e,i}N_{e,i}W(r,h_{i})
\end{equation}
where $dA$ is the projected area orthogonal to the line of sight $dl$, $T_{e,i}$ and $N_{e,i}$ are, respectively, the temperature and the number of electrons of the $i$th gas particle, $W(r,h_i)$ is the SPH kernel adopted in the hydrodynamical simulations to smear out the $y$ signal from each gas particle and $h_i$ is the gas smoothing length.
For the morphological analysis that we perform in this work, we need to compare mock $y$-maps that are representative, in terms of features in the maps and then of S/N, of the real \planck $y$-maps. Therefore, the mock maps are convolved with a Gaussian kernel with $10^{\prime}$ FWHM and gridded in pixels of $1.7^{\prime}$ in size, to mimic the angular resolution of the real maps. Then, the \planck instrumental noise is also taken into account, following a procedure similar to the one described in \citet{Ruppin2019}. A full-sky noise map is realized by using the \planck noise power spectrum and the full-sky map of the standard deviation of the Compton parameter \citep{Planck_y_maps}. From this noise map, cut-outs are extracted randomly, considering the position of the galaxy clusters in the PSZ2 catalogue and then added to the mock $y$-maps to generate \planck-like $y$-maps. Note that point source contamination is not included in these mock $y$-maps. Additional details on the realization of the \planck-like maps are reported in \citet{deAndres2022}, in which these maps are used to infer galaxy clusters masses with a deep learning approach.
Finally, in each of the 4 redshift snapshots selected above we extract the \planck-like maps with S/N $\in$ [$\min$(S/N)$_{Planck}$, $\max$(S/N)$_{Planck}$]. Thus, the number of \thethree clusters selected in each bin are, respectively, 283, 209, 434 and 295 (see the grey dots in Fig.~\ref{fig:M_z}).
As for the real \planck $y$-maps, we use a circular aperture of radius $\theta_{500}$ to define the region to be modelled.

\subsubsection{Dynamical state of \thethree clusters}
\label{sec:The300_dynamical_state}
Being a set of synthetic clusters, for \thethree we can define the dynamical state by taking into account the 3D information from the simulations \citep{Cui2017,Cui2018,Haggar2020,Zhang2022}. In particular, a single parameter, $\chi$, which combines different indicators of the dynamical state, has been defined and already used to segregate different populations in galaxy clusters samples \citep[see e.g. C21,][]{Haggar2020,DeLuca2021}. This study is focused on $R_{500}$ apertures therefore we refer to the definition of $\chi$ in \citet{DeLuca2021}, also used in C21, where only two indicators are adopted in the following way:
\begin{equation}
    \label{eq:chi}
    \chi = \sqrt{ \frac{2}{ \big(\frac{f_s}{0.1}\big)^2 + \big(\frac{\Delta_r}{0.1}\big)^2} }
\end{equation}
where
\begin{equation}
    \label{eq:f_s}
    f_s = \frac{\sum_i M_i}{M_{500}}
\end{equation}
is the total subhalo mass fraction, in which $M_{500}$ is the mass of the cluster computed in a volume of radius equal to $R_{500}$ and $M_i$ are all the subhalo masses in the same volume, and
\begin{equation}
    \label{eq:delta_r}
    \Delta_r = \frac{|\mathbf{r}_{cm} - \mathbf{r}_c|}{R_{500}}
\end{equation}
is the centre-of-mass offset, in which $\mathbf{r}_{cm}$ is the centre-of-mass position of the cluster and $\mathbf{r}_c$ is the centre of the cluster identified with the highest density peak.
Following the suggestions in \citet{DeLuca2021}, relaxed and disturbed systems can be identified, respectively, with $\log_{10}\chi > 0$ and $\log_{10}\chi < 0$.

\section{Methods}
\label{sec:methods}
To study the morphology of the $y$-maps of the clusters, we employ the method developed in C21. This method involves modelling the maps using ZPs and deriving a single parameter from the fitting process to quantify the differences between the maps. We detail this procedure below, after describing the main properties of ZPs.

\subsection{Zernike polynomials}
\label{sec:ZPs}
The ZPs \citep{Zernike1934} are a complete basis of orthogonal functions defined on a unit disc. 
They are widely used to describe aberrations in optical systems \citep[see e.g.][]{Mahajan2007} and a common application is in adaptive optics, to model aberrated wavefronts from atmospheric turbulence \citep[see e.g.][]{Noll1976,Rigaut1991}. More generally, orthogonality and completeness make these polynomials very useful to fit and analyse functions on circular domains. Indeed, they are also applied in a variety of research fields as pattern descriptors \citep[see e.g.][for a review]{Niu2022}. 
We refer to the analytical definition of ZPs given in \citet{Noll1976}:
\begin{equation}
	\label{eq:ZPs}
	\begin{cases}
    Z^m_n(\rho,\theta) = N^m_n R^m_n(\rho) \cos{(m\theta)}\\
    Z^{-m}_n(\rho,\theta) = N^m_n R^m_n(\rho) \sin{(m\theta)}
    \end{cases}
\end{equation}
where $n$ is the polynomial order ($n\in\mathbb{N}$, $n\geqslant0$), $m$ is the azimuthal frequency ($m\in\mathbb{N}$, $m\leqslant n$, $n-m=$ even), $\rho$ and $\theta$ are the polar coordinates ($0\leqslant\rho\leqslant1$, $0\leqslant\theta\leqslant2\pi$), $N^m_n=\sqrt{\frac{2(n+1)}{1+\delta_{m0}}}$ is a normalization factor and
\begin{equation}
    \label{eq:R}
    R^m_n(\rho)=\sum_{s=0}^{(n-m)/2} \frac{(-1)^s (n-s)!}{s!\Bigl(\frac{n+m}{2}-s\Bigr)!\Bigl(\frac{n-m}{2}-s\Bigr)!} \rho^{n-2s}
\end{equation}
is the radial function.
The orthogonality property is given by 
\begin{equation}
    \label{eq:orto}
    \int_0^1 \int_0^{2\pi} Z_n^m(\rho,\theta) Z_{n^{\prime}}^{m^{\prime}}(\rho,\theta) \rho d\rho d\theta = \pi \delta_{nn^{\prime}} \delta_{mm^{\prime}}
    \end{equation}

The $y$ parameter in cluster maps can be modelled as a weighted sum of ZPs as in the following:
\begin{equation}
    \label{eq:y_fit}
    y(\rho,\theta)=\sum_{n} \sum_{m=0}^n c_{n,\pm m} Z_n^{\pm m}(\rho,\theta)
    \end{equation}
where $c_{n,\pm m}$ are the expansion coefficients that quantify the weight of each polynomial in the fit, while the maximum polynomial order $n$ used in the expansion is chosen according to the desired accuracy for modelling the map. About that, we emphasize that the goal of this work is to recognize the morphological regularity of the maps, or rather, to verify if there are inhomogeneities in the distribution of the signal (e.g. clumpy or multi-peaks distributions). This does not require a modelling of the maps in all their details, as already emphasized in C21. In addition, to finalize the modelling we have to take into account the characteristics of our sample in terms of angular resolution in the maps and of the residual noise contamination.
We refer to the study in \citet{Svechnikov2015} and the discussion in Appendix A in C21 to select a suitable number of ZPs for our analysis.
The resolving capability of a Zernike expansion of order $n$, in terms of spatial frequencies $k$ and in units of the inverse radius of the circular aperture, can be roughly derived from $k \approx (n+1)/2\pi$, as described in \citet{Svechnikov2015}.
We emphasize that this is a simple criterion to have an estimate of the spatial resolution that can be reached by using a set of ZPs. However, as showed in Appendix A in C21 by considering the power spectra of the $y$-maps and of the respective Zernike fitting maps, the agreement between modelling and data, for the morphological analysis we are interested in, is still reasonable beyond the value of $k$ derived from the approximation above. In particular, by using a maximum polynomial order $n=8$ the spatial resolution estimated in C21 was $\sim 0.5\theta_{500}$. In \citet{Svechnikov2015} an empirical criterion is derived to also estimate the spatial resolution of a Zernike modelling obtained when changing the maximum order $n$ of the expansion: an increase (decrease) of $n$ of a certain factor corresponds to an increase (decrease) of the spatial resolution of the same factor.
The \planck $y$-maps that we analyse here have a low angular resolution ($10^\prime$, see Section~\ref{sec:PSZ2_cosmo}) and the internal regions of radius $<\theta_{500}$ are poorly resolved therefore, we want to limit the modelling on the scale of the map dimension, i.e. $2\theta_{500}$.
We need a spatial resolution $\sim4$ times lower with respect to the resolution estimated in C21 and we have to reduce the maximum order $n$ of the expansion of the same factor. Thus, we model the \planck $y$-maps with only 6 ZPs, up to the order $n=2$. We show in Section~\ref{sec:results_300} that this choice also allows us to be poor sensitive to the residual noise in the maps.

\subsection{The morphological parameter $\mathcal{C}$}
\label{sec:C_param}
From the Zernike fit we derive a parameter, $\mathcal{C}$, given by
\begin{equation}
    \label{eq:C}
    \mathcal{C} = \sum_{n,m\neq0} |c_{n,\pm m}|^{1/2}
\end{equation}
where $c_{n,\pm m}$ are the Zernike expansion coefficients defined in Eq.~(\ref{eq:y_fit}) and the sum, as explained before, here is truncated to $n=2$. The coefficients $c_{n,\pm m}$ are derived from Eq.~(\ref{eq:y_fit}) by exploiting the orthogonality of the polynomials as expressed in Eq.~(\ref{eq:orto}). They are normalized to the area of the circular aperture, hence computed from
\begin{equation}
    \label{eq:c_nm}
    c_{n,\pm m} = \frac{\sum (y \times Z_n^{\pm m})} {\pi R_{500,px}^2}
\end{equation}
where $R_{500,px}$ is the radius expressed in pixels and the sum is extended to all the pixels. The fit procedure is implemented using the Python package {\small POPPY}\footnote{
https://pypi.org/project/poppy/} \citep{Perrin2012} for calculating ZPs.
By moving from the minimum to the maximum value of $\mathcal{C}$ the maps are recognized as increasingly irregular. 
Note that in the definition in Eq.~(\ref{eq:C}) the coefficients related to ZPs with $m=0$ are neglected. Indeed, in C21 it is shown that the overall contribution of those polynomials is almost invariant when modelling different maps in a wide range of morphologies, then they are not able to distinguish morphological details needed for an accurate classification. On the contrary, the weight of ZPs with $m\neq0$ is negligible when modelling regular (mostly circular) distributions in the maps and it increases in case of complex patterns involving, for example, asymmetries or substructures (see Fig.~\ref{fig:y_maps}). This behaviour is clarified with some examples in the next section.

\section{Results}
\label{sec:results}
In this section we report the results of the Zernike fit applied both on the \planck $y$-maps and on the mock \planck-like $y$-maps. Then, we describe how to convert the morphological analysis derived from the fit in a dynamical-state evaluation.

\subsection{Analysis of the \planck sample}
\label{sec:results_planck}
\begin{figure}
	\includegraphics[width=\columnwidth]{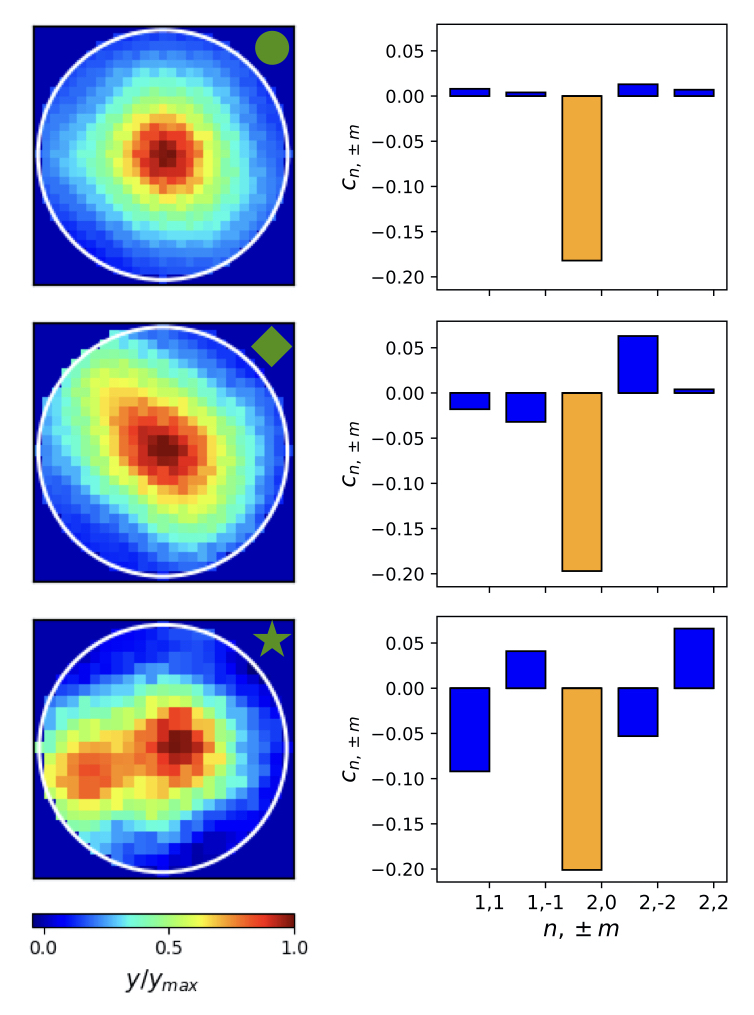}
    \caption{Left: examples of \planck $y$-maps of 3 clusters with different values of the $\mathcal{C}$ parameter. Top: cluster PSZ2 G075.71+13.51 (PSZ2 index 322). Middle: cluster PSZ2 G340.88-33.36 (PSZ2 index 1591). Bottom: cluster PSZ2 G093.94-38.82 (PSZ2 index 430). Each map is centred on the cluster position reported in the PSZ2 catalogue, with side-length equal to $2\theta_{500}$. The white circle is the aperture defined for the Zernike fit. The markers in green in the top-right corner of each map are used to indicate the respective values of the $\mathcal{C}$ parameter in Fig.~\ref{fig:cumulative}.
    Right: bar charts of the $c_{n,\pm m}$ coefficients (see Eq.~(\ref{eq:c_nm})) resulting from the Zernike fit applied on each map on the left. The ZPs on the $x$ axis are ordered following the Noll's scheme \citep{Noll1976}. The polynomial order $n$ increases from left to right. Orange and blue bars refer, respectively, to ZPs with $m=0$ and $m\neq0$.}
    \label{fig:y_maps}
\end{figure}

\begin{figure}
	\includegraphics[width=\columnwidth]{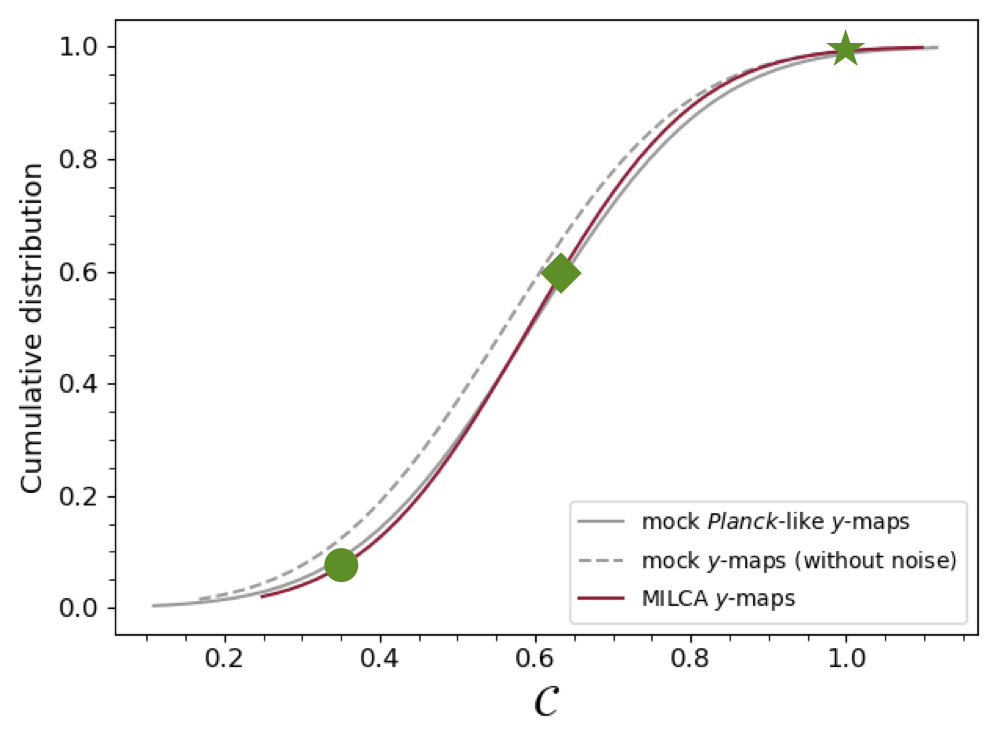}
    \caption{Cumulative distribution of the number of clusters along the $\mathcal{C}$ parameter. The red line refers to the \pszcosmo sample, the grey lines refer to the sample selected from \thethree and considering, respectively, their \planck-like maps (solid line), and their maps without noise (dashed line). The green markers are used to indicate the value of $\mathcal{C}$ for the three maps shown in Fig.~\ref{fig:y_maps}.}
    \label{fig:cumulative}
\end{figure}

We apply the Zernike fit on the $y$-maps of the \pszcosmo sample described in Section~\ref{sec:PSZ2_cosmo}.
We show in Fig.~\ref{fig:y_maps} three examples of \planck $y$-maps with the results of the respective Zernike fits.
In the left panels there are the $y$-maps centred on the clusters coordinates, with the solid white line defining the circular aperture of radius equal to $\theta_{500}$ for each cluster. In the right-side of the figure there are the bar charts of the coefficients $c_{n,\pm m}$ of each ZP used to fit the maps. The ZPs are ordered as in \citet{Noll1976} (see also Fig.~1 in C21), with the order $n$ increasing when moving from left to right, and they are grouped for the azimuthal frequency ($m=0$ in orange and $m\neq0$ in blue). Note that the coefficient $c_{00}$ related to the first polynomial (i.e. $Z_0^0=1$) is neglected because it is just equal to the mean value of $y$ inside the circular aperture. In the top-left panel there is an example of map with a clear circular symmetry: the contribution of ZPs with $m\neq0$ is minimal. The map in the middle-left panel shows an overall prolate distribution but with no resolved substructures and in the fit the weight of ZPs with $m\neq0$ is larger. Instead, in the third map (bottom-left) there is evidence of a possible substructure with respect to the central target and all the ZPs with $m\neq0$ have a non negligible value. As already described, we stress that the coefficients of ZPs with $m=0$ have small differences between the  fits. 
In Fig.~\ref{fig:cumulative} we show the cumulative distribution of the number of clusters along $\mathcal{C}$ for the full sample of clusters. The median value of the distribution is $\mathcal{C}=0.58$, while the 16th and 84th percentiles are at $\mathcal{C}=0.43$ and 0.76, respectively. For the convenience of the reader, in Fig.~\ref{fig:cumulative} we also indicate the value of the $\mathcal{C}$ parameter for the three maps shown in Fig.~\ref{fig:y_maps}, by using the following markers: a circle for the most regular map (top) that has $\mathcal{C}=0.35$, a diamond for the intermediate case (middle) with $\mathcal{C}=0.63$ and a star for the irregular map (bottom) with $\mathcal{C}=0.99$.
The values of $\mathcal{C}$ for each cluster in \pszcosmo are reported in Table~\ref{tab:values}.

It is well known that the study of the morphology of cluster maps, even if an useful method to attempt a cluster dynamical classification, has some limits. In fact, projection effects, limited angular resolution observations and noise contamination in the maps can affect the conclusions on the real state of the clusters. Therefore, it is also important to quantify the efficiency of the morphological analysis in inferring information on the dynamical state.
To do this, here we use the synthetic data set described in Section~\ref{sec:The300}, by exploiting the a priori knowledge of the dynamical state derived from the simulations.

\subsection{Analysis of \thethree sample}
\label{sec:results_300}
\begin{figure}
    \includegraphics[width=\columnwidth]{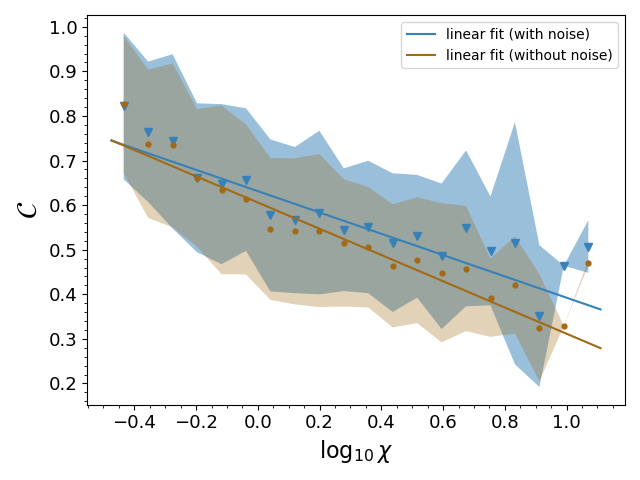}
    \caption{$\mathcal{C}$ versus $\log_{10}\chi$. In blue, the results for the $\mathcal{C}$ parameter computed on the mock \planck-like $y$-maps. In orange, the results for the $\mathcal{C}_{no\_noise}$ parameter computed on the mock $y$-maps generated without noise contamination (see Section~\ref{sec:The300_y_maps}). The blue triangles and the orange points are the mean values of $\mathcal{C}$ and $\mathcal{C}_{no\_noise}$, respectively, in each bin of $\log_{10}\chi$. The shaded areas are at $\pm1\sigma$. The solid blue line is the best linear fit of equation $\mathcal{C} = (-0.24\pm0.02) \log_{10}\chi + (0.63\pm0.01)$. The solid orange line is the best linear fit of equation $\mathcal{C}_{no\_noise} = (-0.29\pm0.02) \log_{10}\chi + (0.61\pm0.01)$.}
    \label{fig:linear_fit}
\end{figure}

In Section~\ref{sec:The300} we describe the sample of clusters selected from \thethree to reproduce the \pszcosmo (see Fig.~\ref{fig:M_z}). We model their \planck-like $y$-maps with the ZPs in the same way as done for the real \textit{Plank} $y$-maps. In Fig.~\ref{fig:cumulative} we show the cumulative distribution of \thethree clusters along $\mathcal{C}$, with the solid line in grey, and we compare it with the cumulative of the \pszcosmo. It is evident that the two distributions are compatible. We confirm this with a two-sample Kolmogorov-Smirnov (KS) test. The KS test is a non-parametric method to quantify the differences between the cumulative distributions of two samples. It provides the KS statistic, which is a measure of the maximum deviation between the two cumulatives, and the $p$-value of the null hypothesis that the two distributions are identical. In our case, KS statistic $=7.1\times10^{-2}$ and $p$-value = 0.67. Thus, we cannot reject the null hypothesis and this means that the mock $y$-maps are well representative of the \planck $y$-maps, i.e. they show the same morphology.
This allows us to pose some questions:
is this morphological classification representative of the dynamical state of the clusters? Or better, is the morphology deduced from $y$-maps correlated to the dynamical state defined from 3D information?
To answer to these questions we study the correlation between the $\mathcal{C}$ parameter and the $\chi$ indicator (we consider its $\log_{10}$) for \thethree.
We perform $10^4$ resamplings of the data set with a bootstrap method and we estimate the mean Pearson correlation coefficient $<r>=-0.38 \pm 0.03$.
In Fig.~\ref{fig:linear_fit}, in blue, we show $\mathcal{C}$ along $\log_{10}\chi$, binning in $\log_{10}\chi$. The triangles are the mean value of $\mathcal{C}$ in each bin of $\log_{10}\chi$ and the shaded area is at $\pm1\sigma$. The solid line in blue is the best linear fit
\begin{equation}
    \mathcal{C} = a \log_{10}\chi + b
    \label{eq:linear_fit}
\end{equation}
where $a=-0.24\pm0.02$ and $b=0.63\pm0.01$. Note that the fit is done by considering the entire sample and not only the mean values in the $\log_{10}\chi$ bins. 

As said before, the noise in the maps is a limiting factor when studying the morphology. Therefore, it is also interesting to see what is its impact on our analysis. Here, we also exploit the availability of mock $y$-maps realized for \thethree with the same angular resolution of \planck but without noise contamination. We refer to this 'clean' data set with a 'no\_noise' subscript. We apply the Zernike fit on these maps as well, following the same procedure above.
For comparison, Fig.~\ref{fig:cumulative} also shows the cumulative distribution of \thethree clusters in this case (see the dashed grey line). It is evident that the impact of the noise is to increase the value of the $\mathcal{C}$ parameter, with this effect being more pronounced in the left tails, i.e. for the most regular maps.
Again, we estimate the correlation $\mathcal{C}_{no\_noise}$ \textit{versus} $\log_{10}\chi$ with a bootstrap method, obtaining $<r_{no\_noise}>=-0.47 \pm 0.02$.  
In Fig.~\ref{fig:linear_fit}, in orange, we show $\mathcal{C}_{no\_noise}$ along $\log_{10}\chi$. As for the previous case, the points are the mean value of $\mathcal{C}_{no\_noise}$ in each bin of $\log_{10}\chi$ and the shaded area is at $\pm1\sigma$. The solid orange line is the best linear fit of Eq.~\ref{eq:linear_fit}, in which the best-fitting parameters are: $a=-0.29\pm0.02$ and $b=0.61\pm0.01$.
As expected, when neglecting the noise contamination in the maps the correlation between the inferred 2D morphology and the 3D dynamical state is higher. However, we notice that the two linear fits shown in Fig.~\ref{fig:linear_fit} are compatible within $2\sigma$. This means that the $\mathcal{C}-\log_{10}\chi$ relation remains poor sensitive to the noise features in the maps.

\subsection{Dynamical state of \planck clusters}
\label{sec:dynamical_state_planck}
\begin{figure}
	\includegraphics[width=\columnwidth]{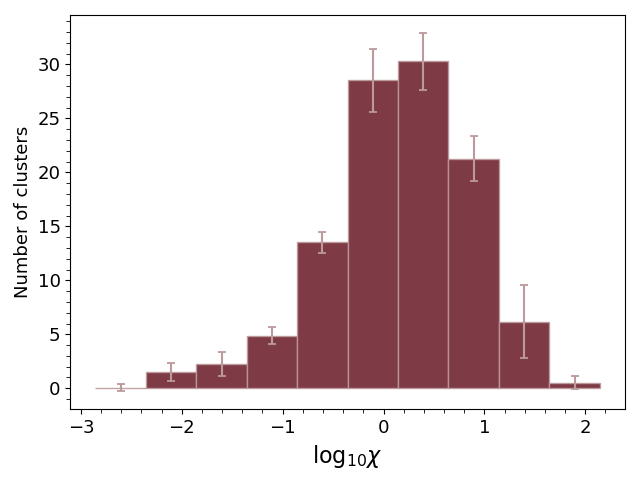}
    \caption{Distribution of the number of \planck clusters along $\log_{10}\chi$. The height of each bar is the mean number of clusters in the respective bin of $\log_{10}\chi$ and the error bar on the top is the standard deviation.}
    \label{fig:chi_planck}
\end{figure}

In the previous section we verified that the sample of mock \planck-like $y$-maps generated for \thethree clusters is well representative, in terms of morphology, of the real \planck $y$-maps. The $\mathcal{C}$ parameter estimated on these maps from the Zernike fit has a moderate linear correlation with the 3D dynamical indicator $\chi$. Then, the best linear fit resulting from this correlation (see Eq.~\ref{eq:linear_fit}) can be used to deduce the relaxation parameter $\chi$, i.e. a proper dynamical state distribution, for the \pszcosmo sample.
For each \planck cluster, we solve the Eq.~\ref{eq:linear_fit} for $\log_{10}\chi$, using the best fit parameters $a$ and $b$ estimated before. We apply a Monte-Carlo error propagation on these latter to compute the error on $\log_{10}\chi$, using $10^4$ random extractions from a 2D Gaussian distribution, accounting for the covariance between $a$ and $b$. 
For each extraction, we compute the respective $\log_{10}\chi$ for each cluster. This generates a distribution of $\log_{10}\chi$ values for each cluster, from which we derive the median as well as the 16th and 84th percentiles. These values are reported in Table~\ref{tab:values}.
We also construct the histogram of the number of \planck clusters along $\log_{10}\chi$ for each random extraction. This is shown in Fig.~\ref{fig:chi_planck}. In each bin of $\log_{10}\chi$ we compute the mean number of clusters, i.e. the height of each bar and the standard deviation, i.e. the error bar on the top.
As described in Section~\ref{sec:The300_dynamical_state}, we apply the threshold $\log_{10}\chi > 0$ to recognize the relaxed clusters.
By using the values reported in Table~\ref{tab:values} we estimate that the fraction of relaxed clusters in the \pszcosmo sample is $61 - 66 \%$.

\begin{figure}
	\includegraphics[width=\columnwidth]{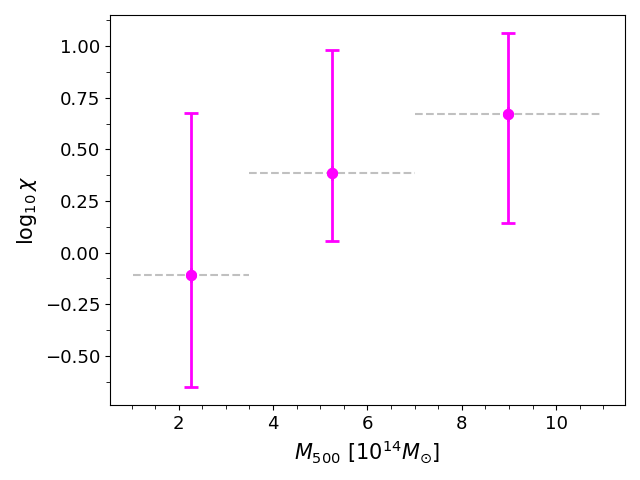}
    \caption{Distribution of the dynamical state indicator, $log_{10}\chi$, as a function of cluster mass, $M_{500}$. The grey dashed lines indicate the mass bins considered, while the pink dots mark the median value of $log_{10}\chi$ in each bin, with error bars corresponding to the 16th and 84th percentiles.}
    \label{fig:chi_M}
\end{figure}

We also analyse the distribution of $\log_{10}\chi$ values as a function of cluster mass, as shown in Fig.~\ref{fig:chi_M}. The grey dashed lines indicate the mass bins considered and the pink dots represent the median of $\log_{10}\chi$ in each bin, with error bars corresponding to the 16th and 84th percentiles. We notice a mild correlation, confirmed by computing the Pearson correlation coefficient ($r=0.41$, $p$-value$=1.13\times 10^{-5}$) and the Spearman correlation coefficient ($\rho=0.49$, $p$-value$=5.38\times 10^{-8}$). The correlation of the dynamical state of the clusters with their mass is not firmly established in the literature.
Various studies, based on both observational data and simulations, have reported differing results. However, we note that some previous works analysing SZ samples found some indications that align with our findings. For instance, \citet{Rossetti2017} found a correlation, even if not strongly confirmed, between mass and cool-core fraction in a sample of clusters from the first release of the \planck catalogue \citep[PSZ1][]{Planck2014}. They found that the high-mass ($>6.5\times 10^{14} M_{\odot}$) subsample features a higher fraction of cool-core clusters compared to the low-mass subsample, also noting that this was consistent with \citet{Mantz2015}, who observed an increasing fraction of relaxed object with the temperature in SZ samples from \planck and SPT. However, both studies emphasized that the statistical significance of these results should be improved by considering larger samples. In contrast, \citet{Lovisari2017} and \citet{Bartalucci2019}, who also analysed samples derived from \planck, did not find any evidence of a correlation between dynamical state and mass. Similarly, \citet{Campitiello2022} confirmed these results in their analysis of \planck clusters in the CHEX-MATE project \citep{CHEX-MATE}. However, upon examining Figure 2 of their study--and as noted by the authors themselves--after a visual classification of cluster maps, it appears that a larger fraction of relaxed object is found at low redshift ($z<0.25$) and with masses $>4\times 10^{14} M_{\odot}$. On the other hand, simulation-based studies have reported opposite results, with evidence of higher fraction of relaxed objects among low-mass clusters \citep[see e.g.][]{Bohringer2010,Fakhouri2010}. We also confirm this using our \thethree sample, where we find a weaker ($\sim30\%$) but opposite correlation between the $\log_{10}\chi$ values derived from the simulations and the masses. In this case, the more massive clusters appear more disturbed, consistent with previous results from \citet{Cui2018} and \citet{Santoni2024}, who also analysed samples from \thethree.
These discrepancies between observations and simulations may be influenced by different biases, such as observational selection effects or the physical assumptions made in the simulations.
We remind that our sample consists of a small number of nearby clusters, covering a mass range of $\sim 1-11 \times 10^{14} M_{\odot}$, therefore it is necessary to extend it to confirm or not a correlation between dynamical state and mass.

\section{Comparison of the dynamical state classification with other works}
\label{sec:comparisons}
The segregation of \planck clusters in terms of the dynamical state was already investigated using different morphological indicators based on multi-wavelength images. Each indicator refers to specific properties of cluster's components, such as gas or galaxies distribution, at different spatial scales. The net result is that the final definition of the state may not be unique.
We believe that a fair comparison of the dynamical state definition of \planck clusters, inferred by standard indicators, with ZPs classification is useful to confirm, or not, their identification.
Therefore, we compare our results with previous works that already analysed subsamples of galaxy clusters from the \planck catalogue.
As pointed out in the literature, in the study of the dynamical state of galaxy clusters there is often no agreement on either the definition of the different classes into which the clusters are grouped (e.g. relaxed, most relaxed, disturbed, most disturbed, mixed, hybrid and so on) or on the adopted parameters and thresholds for the classification.
Also, the region used to study the clusters state (e.g. within $R_{500}$, $R_{200}$ or fraction of these) plays an important role in these analyses \citep{DeLuca2021}. 
When adopting some of these criteria, several approaches are used to quantify their effectiveness. For example, estimating the completeness and/or the purity of the sample \citep{Rasia2013,Lovisari2017,DeLuca2021} or performing statistical tests to identify overlap between the distributions of the dynamical classes \citep{Cialone2018,Lopes2018,DeLuca2021} can help in characterizing the capabilities of the different classifications.
It is also known that, in general, it is more simple to identify relaxed systems with respect to disturbed ones \citep{Lovisari2017}, e.g. due to projection effects. 
Here, we follow the approach in \citet{Lopes2018} to compare our results with the literature. Then, we compute the fraction of agreement, for relaxed clusters, between our dynamical classification and other works. Specifically, we compute the ratio between the number of clusters classified as relaxed both with $\log_{10}\chi$ and a specific parameter and the total number of relaxed clusters recognized with this latter. We report the results in Table~\ref{tab:comparison}. In the following we describe in more details the parameters used for the comparison and also the definition of 'relaxed cluster' adopted in each work.

\citet{Rossetti2016} analysed the cosmology sample of clusters defined from the PSZ1 catalogue \citep{Planck2014}. As the sample that we are taking into account in this work, this first sample was defined to maximize the completeness and the purity of the catalogue for cosmological analyses. The authors analysed \textit{Chandra}, preferentially, and \textit{XMM-Newton} observations for the selected clusters, and they used the offset between the BCG and the X-ray peak as indicator of the dynamical state. An offset less than $0.02R_{500}$ was used to identify relaxed clusters. We find 42 clusters in common with this sample and we estimate a fraction of agreement of 93-96\%. The full sample in \cite{Rossetti2016} was extended to larger redshift with respect to our selection, but they also used two halves, at low ($<0.16$) and high ($>0.16$) redshift, to study the possible evolution of the fraction of the relaxed clusters with the redshift. This kind of study is beyond the aim of this paper due to the already mentioned angular resolution limitations, however, we notice that the fraction of relaxed clusters estimated in the low-redshift subsample ($64\%$), that is almost the same redshift range we use here, is comparable with the fraction of relaxed clusters that we find in this work ($61-66 \%$).

\citet{Rossetti2017} report a more detailed analysis done on the PSZ1 cosmology sample, with the aim of studying the cool-core state of the clusters. Even if this is a more specific question with respect to the characterization of the dynamical state we are interested on, it is interesting to analyse the correlation with our results. The presence of a cool-core, in fact, is usually related to a relaxed state of the cluster, despite in some cases a cool-core could be also present in a disturbed system \citep[see e.g.][]{Menanteau2012}. The concentration parameter, $c$, is devoted to recognize cool-core systems by exploring the cluster cores, namely regions within few hundred kpc from the cluster centre. With respect to our method that uses the ZPs within $R_{500}$, it is more sensitive to the internal regions of the clusters. However, in C21 a good correlation ($>80\%$) was found between the $\mathcal{C}$ parameter from the Zernike fit and $c$, highlighting the advantage of this approach, that is to explore large regions but in more details, being also sensitive to small scales by combining several functions. The $c$ parameter was estimated in \citet{Rossetti2017} within 40 and 400 kpc from the cluster centre, by using a threshold of $0.075$ to recognize cool-core ($c>0.075$) and non cool-core ($c<0.075$) clusters. We refer to these two samples as relaxed and disturbed systems, respectively. Then, the fraction of agreement for the relaxed clusters is 86-88\%.

\citet{Lovisari2017} studied the X-ray morphology of the \planck Early Sunyaev-Zel'dovich (ESZ) sample \citep{Planck_ESZ}, by using \textit{XMM-Newton} images. They applied eight morphological parameters to classify the clusters dynamical state, analysing the performances of each parameter in terms of completeness and purity in revealing relaxed systems, using as reference a previous visual classification. Note that for the visual classification they use three classes: relaxed, disturbed and mix. They found that the best parameters to distinguish different dynamical states are the concentration, $c$, and the centroid shift, $w$. Therefore, we compare our results with only this two indicators. Note that in this case, the concentration parameter was computed in a more external region with respect to the application in \citet{Rossetti2017}, i.e. within $0.1R_{500}$ and $R_{500}$. For the comparison with our results, as usual, we need to count the number of relaxed clusters identified by the single parameters in the reference work. Therefore, based on the discussion in \citet{Lovisari2017}, we apply on $c$ and $w$ the thresholds that maximize the completeness for relaxed clusters. These latter are identified with $c>0.15$ and $w<0.021$. We find a fraction of agreement for relaxed clusters of 87-90\% and 88\%, respectively. The authors also highlighted that a combination of different parameters that are sensitive to different scales in the images can improve the morphological classification, as originally demonstrated in \citet{Rasia2013}. This is in agreement with our discussion on the advantage of the multi-function Zernike fit. They also computed the combined parameter $M$ as a function of $c$ and $w$, but they didn't provide the values of $M$ for each cluster. Therefore, we limit our comparison to $c$ and $w$ only.

\citet{Andrade-Santos2017} analysed a sample of clusters from \planck ESZ as well, by using \textit{Chandra} observations, but they focused on the search of cool-core systems. They used four parameters well adapted to this aim: the concentration parameter, $C_{SB4}$, computed in the internal cluster regions within 40 and 400 kpc and a modified version, indicated as $C_{SB}$, in which the regions analysed are scaled based on $R_{500}$ of each cluster, namely regions within $0.15R_{500}$ and $R_{500}$; the cuspiness of the gas density profile, $\delta$, computed at the radius equal to $0.04R_{500}$; the central gas density, $n_{core}$, computed at $0.01R_{500}$. As for the work of \citet{Rossetti2017}, it is interesting to compare the results of our approach, that model the cluster maps within $R_{500}$, with this analysis that takes into account more internal regions in the clusters. In addition, the sample used in \citet{Andrade-Santos2017} was also analysed in another work by \citet{Lopes2018} in which optical data are also used to segregate the clusters population. Therefore, the comparison with our work that uses tSZ maps is useful to draw a more complete picture of the analysis. However, note that we find a lower number of clusters in common with \citet{Lopes2018}, since their analysis is limited in redshift by the optical surveys used, i.e. the Sloan Digital Sky Survey, the 2dF Galaxy Redshift Survey and the 6dF Galaxy Survey. They used six optical estimators of the dynamical state, but they also suggest that the most reliable way to estimate the cluster state is by using the offset between BCG and X-ray centroid or the magnitude gap, $\Delta m_{12}$, between the BCG and the second BCG. Therefore, we limit the comparison to these two parameters. Furthermore, \citet{Lopes2018} applied some statistical tests to derive the best thresholds for the parameters to separate the two populations of relaxed and disturbed systems (the same approach that has been recovered by \citet{Cialone2018} and \citet{DeLuca2021}. In conclusion, they suggest to modify the thresholds used for the four parameters in \citet{Andrade-Santos2017} in order to have the best performance in the classification. We follow these results and in the comparison with the work of \citet{Andrade-Santos2017} we apply the new thresholds suggested for recognizing the relaxed clusters, namely $C_{SB} > 0.26$, $C_{SB4} > 0.055$, $\delta > 0.46$ and 
$n_{core} > 8 \times 10^{-3}$ cm$^{-3}$. The thresholds for the offset and $\Delta m_{12}$ are $<0.01R_{500}$ and $>1.0$, respectively. The fractions of agreement with our work, for the relaxed clusters, are all larger than 74\%.

\citet{Campitiello2022} studied the morphology of the 118 \planck clusters which compose the CHEX-MATE project \citep{CHEX-MATE}, a Multi-Year \textit{XMM-Newton} Heritage Programme aimed at characterizing the statistical properties of the sample of most recent ($0.05 < z < 0.2$ with $2 \times 10^{14} M_{\odot} < M_{500} < 9 \times 10^{14} M_{\odot}$) and most massive ($z < 0.6$ with $M_{500} > 7.25 \times 10^{14} M_{\odot}$) clusters. They identify four parameters that, used singularly or combined together, are able to recognize different dynamical classes: the concentration parameter, $c$, the centroid shift, $w$, and the second and third-order power ratios, $P_{20}$ and $P_{30}$, all computed within $R_{500}$. It is interesting to note that the power ratios are computed using a similar approach to the Zernike fit, by applying a multipole expansion to the surface brightness within a certain aperture. However, as already discussed in C21, the advantage of the Zernike modelling is the simultaneous use of several functions of different orders which, due to their orthogonality, can be easily combined. This allows for tuning the modelling case by case in terms of the number of functions to use, i.e., choosing the maximum order of the expansion based on the spatial scales one wants to recover. It also allows for adjusting the modelling based on the noise contribution in the maps, as discussed in Section~\ref{sec:ZPs}, whereas contamination in the maps is known to be a limiting factor for the computation of the power ratios \citep{Weissmann2013}. In \citet{Campitiello2022} the authors also combined the four parameters above in a single indicator, $M$, which orders the clusters from the most relaxed to the most disturbed. This approach builds on the method introduced by \citet{Rasia2013} and has been employed in other studies, including the aforementioned \citet{Lovisari2017}, as well as \citet{Cialone2018} and \citet{DeLuca2021}. However, the specific parameters combined and the adopted relationships differ across these works. \citet{Campitiello2022} highlight that $M$ provides a ranking of the clusters, i.e. a continuous sequence based on the dynamical state, rather than a strict segregation in different classes. To test the efficiency of this method, they identify the clusters with lowest (highest) values of $M$ and verify if these were classified as relaxed (disturbed) with a previous visual classification that is used as reference. However, they do not provide exacts values of $M$ to identify these systems. They just consider the number of clusters classified as 'most' relaxed (disturbed) with the visual classification and select the same number of systems starting from the minimum (maximum) value of $M$. Then, they verify if for these clusters the two methods are in agreement.
The objects for which the two classifications match are indicated as relaxed or disturbed, with a mixed population within the two classes. Following these results, they provide a final rank of the clusters by ordering from the most relaxed (the first 15) to the most disturbed (the last 25) on the basis of $M$.
Finally, the thresholds that they suggest for the single parameters are based on this strict classification, i.e. they are values that allow to recognize the most relaxed (disturbed) clusters as classified both with $M$ and visually. For the relaxed clusters the thresholds are: $c>0.49$, $w<0.006$, $P_{20}<1.0 \times 10^{-6}$ and $P_{30}<0.4 \times 10^{-7}$. Note that these are values tighter with respect to the values used for the same parameters in the other works discussed above. This can explain the total agreement with our results, at least for the $c$ and $w$, as reported in Table~\ref{tab:comparison}. On the contrary, the power ratios show the lower fraction. In particular with $P_{20}$ there is the lowest mean agreement, while with $P_{30}$ there is a large range between 58 and 100\%. 
For the $M$ parameter we refer to the more general suggestion of considering $M<0$ to identify the fraction of relaxed systems to compare with our work.

\begin{figure*}
    \centering
	\includegraphics[scale=0.33]{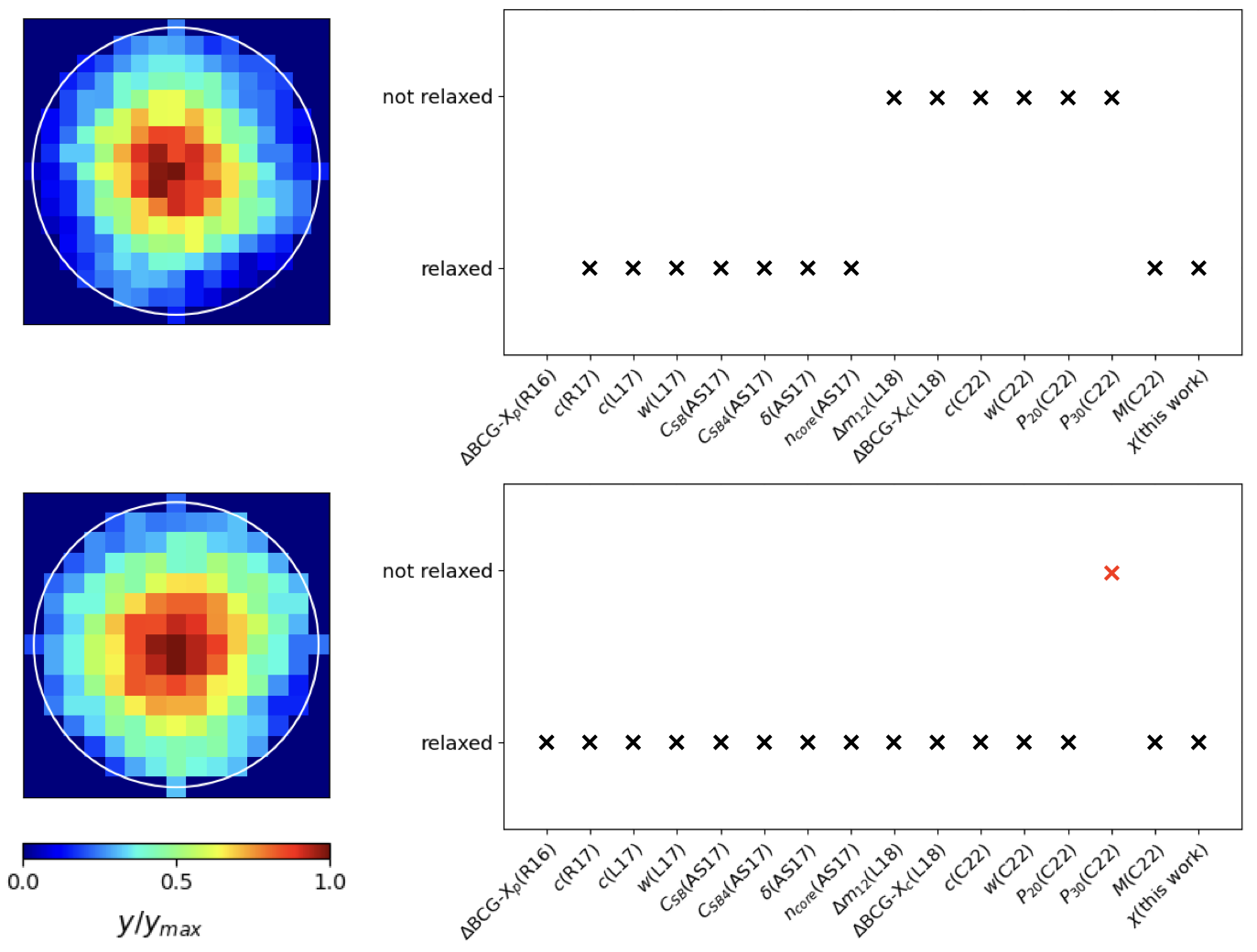}
    \caption{Comparison of the dynamical state classification for PSZ2 G042.81+56.61 (PSZ2 index 156, top) and PSZ2 G056.77+36.32 (PSZ2 index 229, bottom) with the works mentioned in Section~\ref{sec:comparisons}. Left: \textit{Planck} $y$-maps analysed in this work for the two clusters. Right: results of the dynamical state classification based on the papers and the parameters in Table~\ref{tab:comparison} (see Section.~\ref{sec:comparisons} for the thresholds applied to the single parameters to recognize relaxed clusters).}
    \label{fig:comparison}
\end{figure*}

To conclude, we verify that two clusters, PSZ2 G042.81+56.61 and PSZ2 G056.77+36.32, are in common with all the works considered above. Therefore, we check if they are classified in the same way with all the parameters reported in Table~\ref{tab:comparison}. We show the results in Fig.\ref{fig:comparison}. The \planck $y$-maps analysed in this work are shown in the left panels,  while in the plots on the right we indicate how the clusters are classified. We use the 'relaxed' caption to indicate if the cluster satisfies the thresholds reported above for the single parameters, otherwise we refer to the more general class of 'not relaxed'. Indeed, note that some of the other works \citep[see e.g.][]{Lovisari2017,Campitiello2022} also defined an intermediate group of 'mixed' clusters between the relaxed and the disturbed ones. We do not go into details of this more complex classification. With the $\chi$ parameter we classify both the clusters as relaxed ($\log_{10}\chi >0$). 

For PSZ2 G042.81+56.61 (on the top in the figure) our result agrees with the X-ray classification in \citet{Rossetti2016,Rossetti2017,Lovisari2017,Andrade-Santos2017}, but not with the optical analysis in \citet{Lopes2018}. Note that, as already mentioned, for the parameters used in \citet{Andrade-Santos2017} we apply the thresholds suggested in \citet{Lopes2018}. In \citet{Campitiello2022} the cluster does not satisfy the thresholds for the single parameters and it is not in the first 15 clusters that the authors identify as most relaxed. However, $M=-0.05$, therefore we indicate it as relaxed by using the less strict threshold we mentioned above ($M<0$).

PSZ2 G056.77+36.32 (on the bottom in the figure) is classified as relaxed with all the parameters but $P_{30}$ in \citet{Campitiello2022}. However, we verify that the value of this parameter is $P_{30}=1.0^{+0.4}_{-0.6} \times 10^{-7}$, namely the lower limit correspond to the threshold that the authors indicate for the relaxed clusters ($P_{30}<0.4 \times 10^{-7}$). Therefore, we mark the results for this parameter in red in Fig.~\ref{fig:comparison}.

\begin{table*}
  \begin{center}
    \caption{Agreement between the relaxation classification derived from ZPs and other indicators (third column) present in the literature (first column) for common cluster samples (the number of objects in the second column).}
    \label{tab:comparison}
    \begin{tabular}{cccc}
    \hline \hline
      \\
      Reference & Number of common clusters & Morphological indicator & Fraction of agreement \\\\
      \hline\hline
      \citet{Rossetti2016} (R16) & 42 & BCG -- X-ray peak offset & 0.93 - 0.96\\
      \hline
      \citet{Rossetti2017} (R17) & 46 & $c$ & 0.86 - 0.88\\
      \hline
      \multirow{2}{*}{\citet{Lovisari2017} (L17)} &\multirow{2}{*}{19} & $c$ & 0.87 - 0.90\\ 
      & & $w$ & 0.88\\
      \hline
      \multirow{4}{*}{\citet{Andrade-Santos2017} (AS17)} &\multirow{4}{*}{51} & $C_{SB}$ & 0.87 - 0.89\\ 
      & & $C_{SB4}$ & 0.80 - 0.81\\
      & & $\delta$ & 0.76 - 0.81\\       
      & & $n_{core}$ & 0.78 - 0.81\\   
      \hline
      \multirow{2}{*}{\citet{Lopes2018} (L18)} &\multirow{2}{*}{44} & $\Delta m_{12}$ & 0.75 - 0.80\\ 
      & & BCG -- X-ray centroid offset & 0.74 - 0.78\\
      \hline
      \multirow{4}{*}{\citet{Campitiello2022} (C22)} &\multirow{4}{*}{23} & $c$ & 1.0\\ 
      & & $w$ & 1.0\\
      & & $P20$ & 0.60 - 0.75\\       
      & & $P30$ & 0.58 - 1.0\\   
      & & $M$ & 0.80\\   
      \hline
    \end{tabular}
  \end{center}
\end{table*}

\section{Conclusions}
\label{sec:conclusions}
The study of the morphology of galaxy cluster maps is an observational approach that attempts to infer the dynamical state of these systems. 
In this work we conduct a morphological analysis using the Zernike polynomials, chosen for their straightforward analytical properties that are well-suited for modelling circular cluster maps. This novel approach was introduced in \citet{Capalbo2021} for analysing mock high-angular resolution images of hydrodynamical simulated clusters from \theth project. Here, we apply for the first time the Zernike polynomials on real data. 
We analyse Compton parameter maps from the \planck satellite, focusing on galaxy clusters at $z<0.1$ due to the better resolution on mapping nearby systems. In particular, we model the maps within $R_{500}$ for each cluster and we derive a single parameter, $\mathcal{C}$, from the Zernike fit, to describe their morphology.
The number of functions used for modelling must be tuned taking into account both the angular resolution of the maps and the noise spatial contamination. In this case, we choose the number of polynomials in order to be sensitive to the resolving scale of $\sim 2\theta_{500}$.
Mock maps are used to validate the analysis. These are generated for synthetic clusters in \theth project, resembling the real sample in terms of masses, redshift, signal-to-noise ratio and $\mathcal{C}$ parameter. The synthetic clusters were previously classified for their dynamical state using a 3D indicator such as the relaxation parameter, $\chi$. A linear correlation of approximately 38\% is estimated between $\mathcal{C}$ and $\log_{10}\chi$ for the simulated clusters. Note that this is a lower correlation with respect to the results reported in \citet{Capalbo2021} (correlation $>60\%$ at $z=0$), mainly due to the low angular resolution of the \planck maps and the noise contamination. Despite this limitation, we use this correlation to derive a distribution of the dynamical state of the real clusters in terms of $\chi$. Then, a fraction of $\sim63\%$ of the clusters in the \planck sample are recognized as relaxed systems. Preliminary results of this study are presented in \citep{Capalbo2023}.
The dynamical state classification derived in this work is also compared with previous analyses in the literature conducted at different wavelengths and with different indicators. Note that some of those works also analysed the differences, in terms of dynamical state, between samples of clusters derived from SZ and X-ray surveys. This is beyond the aim of this paper and will be addressed in future works after the calibration of the Zernike modelling on X-ray data.
As already mentioned, the proposed approach is subject to strong application limitations, such as where noise and angular resolution are of primary concern as in \planck maps, restricting our analysis to nearby objects only. To address this challenge, we are applying the same method to higher resolution maps such as those from NIKA2 Sunyaev-Zel’dovich Large Program \citep{Perotto2023}. 
The application of Zernike polynomials is also promising on X-ray maps. Enjoying the reduced noise level and carefully choosing the number of polynomials in the modelling, it is possible to extend this method, as preliminary shown in \citet{Capalbo2022}, and as is being done on CHEX-MATE clusters.

\begin{acknowledgements}
The authors are grateful to the anonymous referee for the helpful comments and suggestions, which improved this work.
This work has been made possible by \theth collaboration. The simulations used in this paper have been performed in the MareNostrum Supercomputer at the Barcelona Supercomputing Center, thanks to CPU time granted by the Red Espa\~{n}ola de Supercomputaci\'on. This work has received financial support from the European Union's Horizon 2020 Research and Innovation programme under the Marie Sklodowskaw-Curie grant agreement number 734374, i.e. the LACEGAL project. Some of the results in this paper have been derived using the {\small healpy} and {\small HEALPix} packages. This research made use of {\small POPPY}, an open-source optical propagation Python package originally developed for the James Webb Space Telescope project \citep{Perrin2012}.
VC, MDP and AF acknowledge financial support from PRIN-MIUR grant
20228B938N "Mass and selection biases of galaxy clusters: a multi-probe approach" funded by the European Union Next generation EU, Mission 4 Component 1 CUP C53D2300092 0006 and from Sapienza Università di Roma, thanks to Progetti di Ricerca Medi 2022, RM1221816758ED4E.
AF acknowledges the project "Strengthening the Italian Leadership in ELT and SKA (STILES)", proposal nr. IR0000034, admitted and eligible for funding from the funds referred to in the D.D. prot. no. 245 of August 10, 2022 and D.D. 326 of August 30, 2022, funded under the program "Next Generation EU" of the European Union, “Piano Nazionale di Ripresa e Resilienza” (PNRR) of the Italian Ministry of University and Research (MUR), “Fund for the creation of an integrated system of research and innovation infrastructures”, Action 3.1.1 "Creation of new IR or strengthening of existing IR involved in the Horizon Europe Scientific Excellence objectives and the establishment of networks”.
WC is supported by the Atracci\'{o}n de Talento Contract no. 2020-T1/TIC-19882 granted by the Comunidad de Madrid in Spain and the science research grants from the China Manned Space Project. He also thanks the Ministerio de Ciencia e Innovaci\'on (Spain) for financial support under Project grant PID2021-122603NB-C21, ERC: HORIZON-TMA-MSCA-SE for supporting the LACEGAL-III project with grant number 101086388, and the science research grants from the China Manned Space Project with No. CMS-CSST-2021-A01, and CMS-CSST-2021-A03.
\end{acknowledgements}

\bibliographystyle{aa} 
\bibliography{biblio}

\begin{appendix} 
\onecolumn

\section{Morphological and dynamical state parameters}
The morphological parameter, $\mathcal{C}$, derived from the Zernike fit and the deduced dynamical state indicator, $\log_{10}\chi$, for the selected \planck clusters are reported in Table~\ref{tab:values}. The parameters are calculated within $R_{500}$.

\begin{longtable}{cccccc}
\caption{$\mathcal{C}$ and $\log_{10}\chi$ for the selected 109 \planck clusters at $z<0.1$, listed with index and PSZ2 name.}
\label{tab:values}\\
\hline\hline
\\
Index & PSZ2 Name & $\mathcal{C}$ & \multicolumn{3}{c}{$\log_{10}\chi$ (percentiles)}\\
\cmidrule(lr){4-6}
      &           &               & $16th$ & $50th$ & $84th$ \\  
\\      
\hline\hline
\endfirsthead
\caption{continued.}\\
\hline\hline
\\
Index & PSZ2 Name & $\mathcal{C}$ & \multicolumn{3}{c}{$\log_{10}\chi$ (percentiles)}\\
\cmidrule(lr){4-6}
      &           &               & $16th$ & $50th$ & $84th$ \\
\\
\hline\hline
\endhead
\hline
\endfoot
21 & PSZ2 G006.49+50.56 & 0.48 & 0.55 & 0.61 & 0.68\\ 
22 & PSZ2 G006.68-35.55 & 0.45 & 0.67 & 0.74 & 0.82\\
32 & PSZ2 G008.80-35.18 & 0.79 & -0.74 & -0.67 & -0.60\\
43 & PSZ2 G012.81+49.68 & 0.60 & 0.07 & 0.11 & 0.16\\ 
63 & PSZ2 G019.48-80.97 & 0.46 & 0.65 & 0.72 & 0.79\\ 
83 & PSZ2 G025.13-75.88 & 0.45 & 0.68 & 0.75 & 0.83\\  
99 & PSZ2 G028.63+50.15 & 0.74 & -0.53 & -0.47 & -0.41\\   
102 & PSZ2 G029.06+44.55 & 0.55 & 0.27 & 0.32 & 0.37\\ 
112 & PSZ2 G031.93+78.71 & 0.66 & -0.16 & -0.11 & -0.07\\  
118 & PSZ2 G033.46-48.43 & 0.24 & 1.50 & 1.62 & 1.78\\  
119 & PSZ2 G033.81+77.18 & 0.55 & 0.29 & 0.33 & 0.39\\  
122 & PSZ2 G034.38+51.57 & 0.94 & -1.43 & -1.30 & -1.20\\  
145 & PSZ2 G040.03+74.95 & 0.55 & 0.30 & 0.35 & 0.40\\  
148 & PSZ2 G040.58+77.12 &  0.65 & -0.14 & -0.10 & -0.05\\  
156 & PSZ2 G042.81+56.61 & 0.34 & 1.11 & 1.21 & 1.33\\  
160 & PSZ2 G044.20+48.66 & 0.65 & -0.13 & -0.09 & -0.05\\  
162 & PSZ2 G044.46-65.42 & 0.95 & -1.46 & -1.33 & -1.22\\   
167 & PSZ2 G045.10-43.18 & 0.82 & -0.86 & -0.78 & -0.71\\  
182 & PSZ2 G046.47-49.44 & 0.45 & 0.67 & 0.74 & 0.82\\ 
190 & PSZ2 G048.10+57.16 & 0.51 & 0.45 & 0.50 & 0.56\\  
200 & PSZ2 G049.32+44.37 & 0.76 & -0.61 & -0.55 & -0.49\\ 
201 & PSZ2 G049.69-49.46 & 0.67 & -0.20 & -0.16 & -0.12\\ 
214 & PSZ2 G053.55-29.85 & 0.67 & -0.21 & -0.17 & -0.12\\ 
229 & PSZ2 G056.77+36.32 & 0.40 & 0.87 & 0.95 & 1.05\\ 
235 & PSZ2 G057.61+34.93 & 0.44 & 0.71 & 0.78 & 0.87\\ 
237 & PSZ2 G057.78+52.32 & 1.10 & -2.16 & -1.98 & -1.82\\ 
238 & PSZ2 G057.80+88.00 & 0.47 & 0.61 & 0.67 & 0.75\\ 
239 & PSZ2 G057.92+27.64 & 0.35 & 1.06 & 1.15 & 1.27\\
259 & PSZ2 G062.44-46.43 & 0.53 & 0.36 & 0.41 & 0.47\\ 
261 & PSZ2 G062.94+43.69 & 0.39 & 0.91 & 0.99 & 1.09\\ 
268 & PSZ2 G065.32-64.84 & 0.56 & 0.27 & 0.31 & 0.36\\ 
289 & PSZ2 G068.22+15.18 & 0.69 & -0.31 & -0.26 & -0.21\\ 
322 & PSZ2 G075.71+13.51 & 0.35 & 1.07 & 1.17 & 1.28\\ 
339 & PSZ2 G080.16+57.65 & 0.73 & -0.50 & -0.44 & -0.39\\ 
344 & PSZ2 G080.65-46.81 & 0.69 & -0.30 & -0.25 & -0.21\\ 
349 & PSZ2 G081.31-68.56 & 0.56 & 0.25 & 0.29 & 0.34\\ 
372 & PSZ2 G084.81-62.18 & 0.56 & 0.23 & 0.27 & 0.32\\ 
403 & PSZ2 G089.52+62.34 & 0.76 & -0.62 & -0.55 & -0.49\\ 
425 & PSZ2 G093.42-43.21 & 0.62 & 0.00 & 0.04 & 0.08\\ 
427 & PSZ2 G093.92+34.92 & 0.36 & 1.02 & 1.11 & 1.22\\ 
430 & PSZ2 G093.94-38.82 & 0.99 & -1.65 & -1.51 & -1.39\\
460 & PSZ2 G098.10+30.30 & 0.66 & -0.16 & -0.12 & -0.08\\ 
492 & PSZ2 G101.68-49.21 & 0.59 & 0.11 & 0.15 & 0.19\\ 
495 & PSZ2 G103.40-32.99 & 0.72 & -0.43 & -0.37 & -0.32\\ 
508 & PSZ2 G105.55+77.21 & 0.62 & -0.01 & 0.03 & 0.07\\ 
539 & PSZ2 G110.98+31.73 & 0.36 & 1.02 & 1.11 & 1.22\\
544 & PSZ2 G112.48+56.99 & 0.60 & 0.07 & 0.11 & 0.16\\
547 & PSZ2 G113.02-64.68 & 0.46 & 0.66 & 0.73 & 0.80\\
556 & PSZ2 G114.79-33.71 & 0.35 & 1.07 & 1.17 & 1.29\\ 
562 & PSZ2 G115.25-72.07 & 0.41 & 0.83 & 0.91 & 1.01\\ 
576 & PSZ2 G117.98-55.88 & 0.84 & -0.97 & -0.88 & -0.80\\ 
618 & PSZ2 G125.68-64.12 & 0.61 & 0.05 & 0.09 & 0.13\\ 
670 & PSZ2 G136.64-25.03 & 0.55 & 0.27 & 0.32 & 0.37\\ 
672 & PSZ2 G136.92+59.46 & 0.77 & -0.64 & -0.57 & -0.51\\ 
676 & PSZ2 G137.74-27.08 & 0.93 & -1.36 & -1.24 & -1.14\\ 
695 & PSZ2 G143.00-77.70 & 0.54 & 0.33 & 0.38 & 0.43\\ 
715 & PSZ2 G146.35-15.59 & 0.69 & -0.32 & -0.27 & -0.23\\ 
726 & PSZ2 G149.63-84.19 & 0.61 & 0.05 & 0.10 & 0.14\\ 
770 & PSZ2 G161.39+26.27 & 0.43 & 0.77 & 0.85 & 0.94\\ 
816 & PSZ2 G172.74+65.30 & 0.77 & -0.63 & -0.56 & -0.50\\ 
943 & PSZ2 G209.54-36.50 & 0.39 & 0.93 & 1.02 & 1.12\\ 
986 & PSZ2 G219.45-35.90 & 0.67 & -0.22 & -0.18 & -0.14\\ 
990 & PSZ2 G220.53-38.55 & 0.85 & -1.02 & -0.93 & -0.85\\ 
1019 & PSZ2 G226.16-21.95 & 0.60 & 0.07 & 0.11 & 0.16\\
1041 & PSZ2 G229.93+15.30 & 0.50 & 0.48 & 0.54 & 0.61\\ 
1042 & PSZ2 G230.28-24.42 & 0.89 & -1.17 & -1.06 & -0.97\\ 
1044 & PSZ2 G230.29-47.13 & 0.62 & 0.01 & 0.05 & 0.10\\
1067 & PSZ2 G234.59+73.01 & 0.61 & 0.03 & 0.08 & 0.12\\ 
1078 & PSZ2 G239.29+24.75 & 0.40 & 0.87 & 0.95 & 1.05\\ 
1098 & PSZ2 G242.41-37.40 & 0.64 & -0.07 & -0.03 & 0.01\\ 
1101 & PSZ2 G243.02+42.87 & 1.11 & -2.17 & -1.98 & -1.83\\ 
1104 & PSZ2 G243.64+67.74 & 0.56 & 0.27 & 0.31 & 0.36\\
1119 & PSZ2 G246.50-26.09 & 0.48 & 0.57 & 0.63 & 0.71\\
1155 & PSZ2 G252.99-56.09 & 0.51 & 0.44 & 0.49 & 0.55\\ 
1156 & PSZ2 G253.04+36.83 & 0.69 & -0.28 & -0.24 & -0.19\\ 
1209 & PSZ2 G262.36-25.15 & 0.58 & 0.15 & 0.20 & 0.24\\ 
1215 & PSZ2 G263.19-25.19 & 0.54 & 0.31 & 0.36 & 0.42\\ 
1225 & PSZ2 G265.02-48.96 & 0.43 & 0.76 & 0.84 & 0.92\\ 
1227 & PSZ2 G265.21-24.83 & 0.61 & 0.04 & 0.08 & 0.12\\
1246 & PSZ2 G269.31-49.87 & 0.39 & 0.91 & 1.00 & 1.10\\
1267 & PSZ2 G272.08-40.16 & 0.51 & 0.43 & 0.49 & 0.55\\ 
1272 & PSZ2 G272.88+19.14 & 0.44 & 0.71 & 0.78 & 0.87\\  
1320 & PSZ2 G283.91+73.87 & 0.59 & 0.11 & 0.15 & 0.20\\ 
1346 & PSZ2 G287.46+81.12 & 0.80 & -0.80 & -0.73 & -0.66\\ 
1350 & PSZ2 G287.72+26.46 & 0.77 & -0.63 & -0.57 & -0.51\\
1363 & PSZ2 G290.19+30.16 & 0.55 & 0.28 & 0.33 & 0.38\\ 
1378 & PSZ2 G293.12-70.85 & 0.58 & 0.15 & 0.19 & 0.23\\
1403 & PSZ2 G296.42-32.49 & 0.66 & -0.16 & -0.11 & -0.07\\ 
1429 & PSZ2 G303.75+33.70 & 0.60 & 0.08 & 0.12 & 0.16\\
1431 & PSZ2 G304.44+32.48 & 0.35 & 1.09 & 1.18 & 1.30\\
1435 & PSZ2 G304.91+45.43 & 0.62 & 0.00 & 0.05 & 0.09\\
1450 & PSZ2 G306.66+61.06 & 0.50 & 0.47 & 0.53 & 0.59\\
1451 & PSZ2 G306.77+58.61 & 0.49 & 0.52 & 0.58 & 0.65\\ 
1471 & PSZ2 G311.98+30.71 & 0.56 & 0.27 & 0.31 & 0.36\\ 
1474 & PSZ2 G312.62+35.05 & 0.57 & 0.22 & 0.26 & 0.31\\
1479 & PSZ2 G313.33+30.29 & 0.67 & -0.20 & -0.15 & -0.11\\
1492 & PSZ2 G316.31+28.53 & 0.36 & 1.03 & 1.12 & 1.24\\
1514 & PSZ2 G321.98-47.96 & 0.53 & 0.37 & 0.42 & 0.47\\
1518 & PSZ2 G322.77+59.52 & 0.80 & -0.79 & -0.71 & -0.65\\
1557 & PSZ2 G331.12+62.31 & 0.46 & 0.66 & 0.73 & 0.81\\
1558 & PSZ2 G331.96-45.74 & 0.72 & -0.45 & -0.39 & -0.34\\ 
1560 & PSZ2 G332.23-46.37 & 0.47 & 0.61 & 0.68 & 0.75\\
1567 & PSZ2 G335.58-46.44 & 0.55 & 0.30 & 0.35 & 0.40\\
1569 & PSZ2 G336.60-55.43 & 0.50 & 0.48 & 0.53 & 0.60\\
1570 & PSZ2 G336.95-45.75 & 0.72 & -0.45 & -0.39 & -0.34\\
1591 & PSZ2 G340.88-33.36 & 0.63 & 0.03 & -0.01 & 0.05\\
1604 & PSZ2 G342.82-30.48 & 0.50 & 0.48 & 0.53 & 0.60\\ 
1612 & PSZ2 G345.38-39.32 & 0.70 & -0.32 & -0.28 & -0.23\\
1641 & PSZ2 G356.25-76.07 & 0.29 & 1.31 & 1.42 & 1.56\\
\end{longtable}
\end{appendix}

\end{document}